\documentclass{webofc}
\usepackage[varg]{txfonts}   
\begin{document}
\title{Intriguing feature of multiplicity distributions}
%
%

\author{Maciej Rybczy\'nski\inst{1}\fnsep\thanks{\email{maciej.rybczynski@ujk.edu.pl}} \and
        Grzegorz Wilk\inst{2}\fnsep\thanks{\email{grzegorz.wilk@ncbj.gov.pl}} \and
        Zbigniew W\l odarczyk\inst{1}\fnsep\thanks{\email{zbigniew.wlodarczyk@ujk.kielce.pl}}
}

\institute{Institute of Physics, Jan Kochanowski University, \'Swi\c{e}tokrzyska 15, 25-406 Kielce, Poland
\and
           National Centre for Nuclear Research, Ho\.za 69, 00-681 Warsaw, Poland
          }

\abstract{%
   Multiplicity distributions, $P(N)$, provide valuable information on the mechanism of the production process. We argue that the observed $P(N)$ contain more information (located in the small $N$ region) than expected and used so far. We demonstrate that it can be retrieved by analysing specific combinations of the experimentally measured values of $P(N)$ which we call {\it modified combinants}, $C_j$, and which show distinct oscillatory behavior, not observed in the usual phenomenological forms of the $P(N)$ used to fit data. We discuss the possible sources of these oscillations and their impact on our understanding of the multiparticle production mechanism.
  }
\maketitle

\section{Introduction}
\label{Introduction}

\begin{figure}[t]
\begin{center}
\includegraphics[scale=0.35]{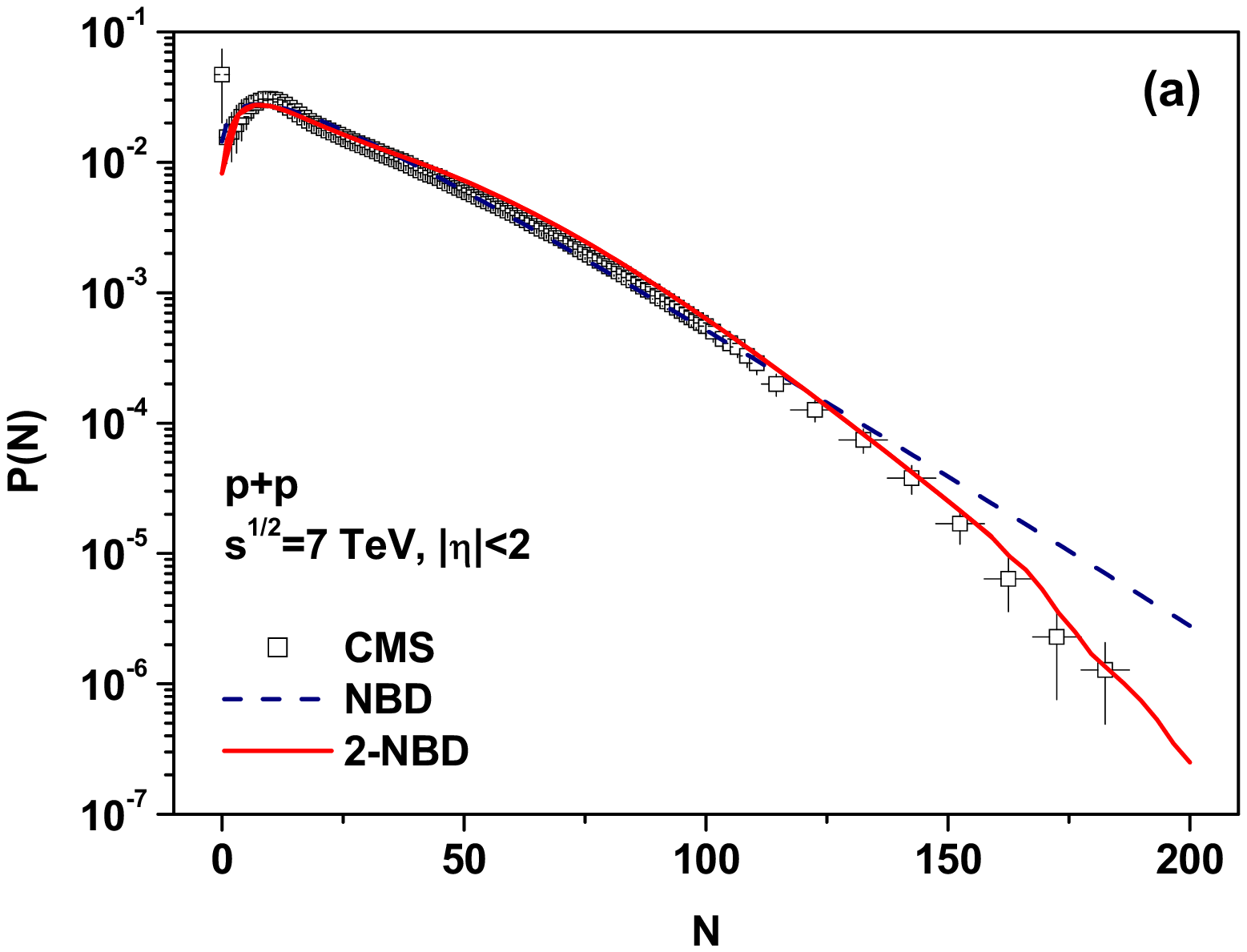}\hspace{10mm}
\includegraphics[scale=0.35]{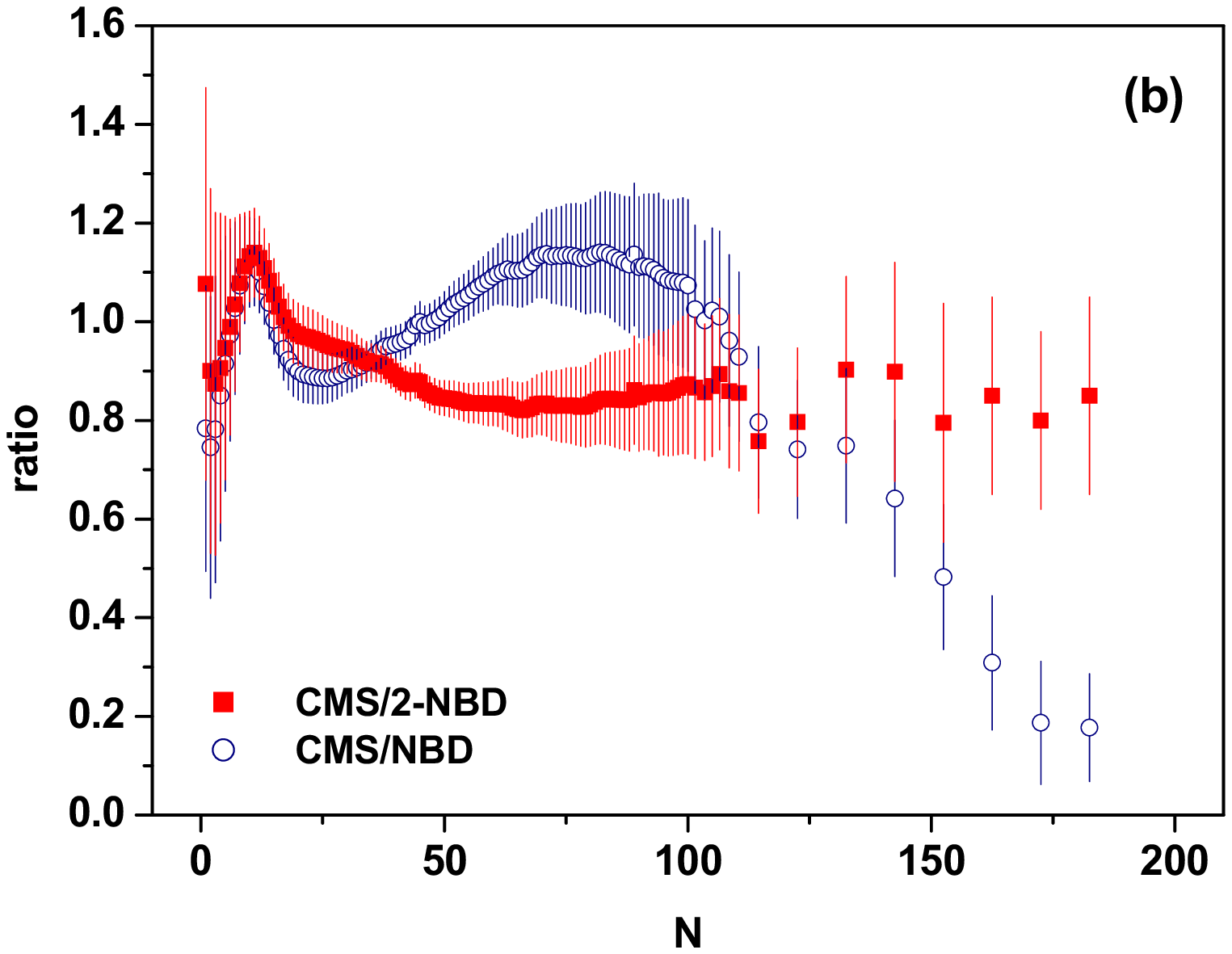}\\
\vspace{-5mm}
\includegraphics[scale=0.35]{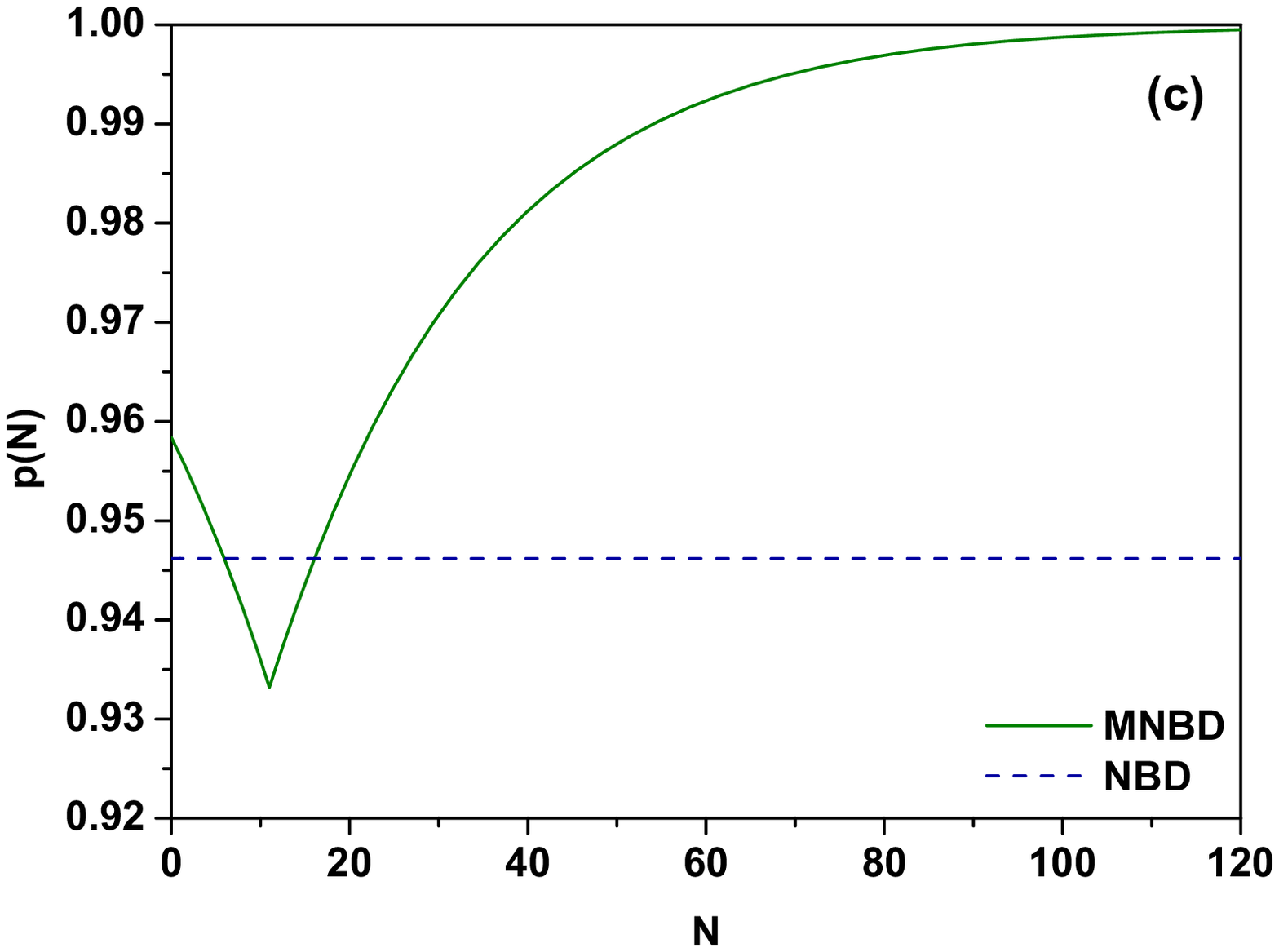}\hspace{10mm}
\includegraphics[scale=0.35]{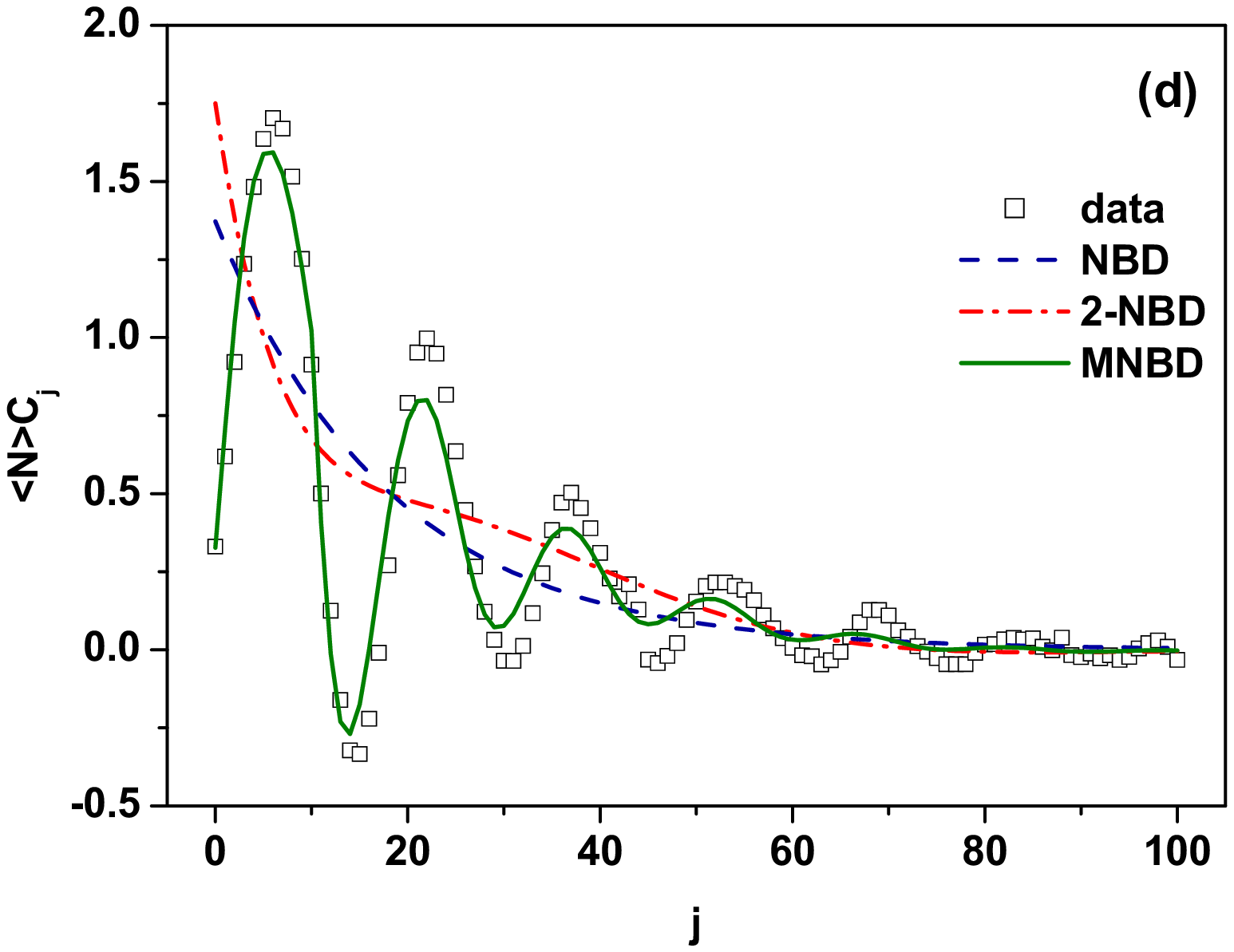}
\end{center}
\vspace{-10mm}
\caption{$(a)$ Charged hadron multiplicity distributions for the pseudorapidity range $|\eta| < 2$ at $\sqrt{s} = 7$ TeV, as given by the CMS experiment \cite{CMS} (points), compared with the NBD for parameters $\langle N\rangle = 25.5$ and $k = 1.45$ (dashed line) and with the $2$-component NBD (solid line) with parameters from \cite{PG}. $(b)$ Multiplicity dependence of the ratio $R = data/fit$ for the NBD (circles) and for the $2$-component NBD for the same data as in panel $(a)$ (squares). $(c)$ The multiplicity dependence of the modified probability of particle emission $p$ in the MNBD as given by Eq. (\ref{Non}) resulting in flat $R=R(N)=1$. $(d)$ Coefficients $C_j$ emerging from the CMS data used in panel $(a)$ compared with the NB and $2$-NBD fits shown there and with the $C_j$ obtained from the MNB with modifications proposed in \cite{JPG} and shown in panel $(c)$.
}
\label{Fig1}
\vspace{-5mm}
\end{figure}

The multiplicity distribution, $P(N)$, is an important characteristic of the  multiparticle  production process, one of the first observables measured in any multiparticle production experiment \cite{Kittel}. The way in which the consecutive $P(N)$ are connected reflects the dynamics of the multiparticle production process. In the simplest case one assumes that  the multiplicity $N$ is directly influenced only by its neighbouring multiplicities $(N \pm 1)$ in the way dictated by the simple recurrence relation:
\begin{equation}
(N+1)P(N+1) = g(N)P(N)\quad {\rm where}\quad g(N) = \alpha + \beta N,\label{rr1}
\end{equation}
The most popular forms of $P(N)$ emerging from this recurrence relation are ($p$ denotes probability of particle emission):
\begin{eqnarray}
\hspace{-15mm}&& P(N) = \frac{K!}{N!(K - N)!} p^N (1 - p)^{K-N},~~~~~\alpha = \frac{Kp}{1-p},~~ \beta = -\frac{\alpha}{K};\qquad {\rm Binomial~Distribution~(BD)};\label{BD}\\
\hspace{-15mm}&& P(N) = \frac{\lambda^N}{N!} \exp( - \lambda),\qquad\qquad ~~~~~~~\quad \alpha = \lambda,~~ \beta = 0;\qquad\quad{\rm Poisson~Distribution~(PD)}; \label{PD}\\
\hspace{-15mm}&& P(N) = \frac{\Gamma(N+k)}{\Gamma(N+1)\Gamma(k)} p^N (1 - p)^k,\quad ~~~\alpha = kp,~~~ \beta = \frac{\alpha}{k};\quad {\rm Negative~Binomial~Distribution~(NBD)}.\label{NBD}
\end{eqnarray}
Usually the first choice of $P(N)$ in fitting data is a single NBD. However, with growing energy and  number of produced secondaries it increasingly deviates from the data for large $N$ (see Fig. \ref{Fig1}) and is therefore replaced either by combinations of two \cite{GU}, three \cite{Z}, or multi-component NBDs \cite{DN}, or by some other forms of $P(N)$ \cite{Kittel,DG,KNO-2,MF,HC}. However, as seen in Fig. \ref{Fig1} $(a)-(b)$, such a procedure only improves the agreement at large $N$, the ratio $R = data/fit$ deviates dramatically from unity at small $N$ for all fits. This observation, when taken seriously, suggests that there must be some additional information hidden in the small $N$ region, not investigated yet \cite{JPG}. In \cite{JPG} we retrieved it using a single NBD form of $P(N)$ in which we allowed for the multiplicity dependence of the particle emission ratio $p = m/(m+k)$ in Eq. (\ref{NBD}). It turns out that for
\begin{equation}
m = m(N) = c\exp\left[ a_1 | N - b| + a_2( N - b)^4\right] \label{Non}
\end{equation}
with parameters $c = 20.252$, $a_1 = 0.044$, $a_2 = 1.04\cdot 10^{-9}$ and $b = 11$, one gets the desired flat behavior of $R$ as a function of multiplicity $N$,  now $R=1$ for all $N$ \cite{JPG}. Such a choice corresponds to a rather complicated, nonlinear and non monotonic spout-like form of $g(N)$ in the recurrence relation  Eq. (\ref{rr1}) and to a non monotonic, depending on multiplicity, probability of particle emission, $p = p(N)$, with a sharp minimum around $N=10$, after which $p(N)$ grows steadily, see Fig. \ref{Fig1} $(c)$.

\section{Modified combinants $C_j$}
\label{ModComb-d}

\begin{figure}[t]
\begin{center}
\includegraphics[scale=0.35]{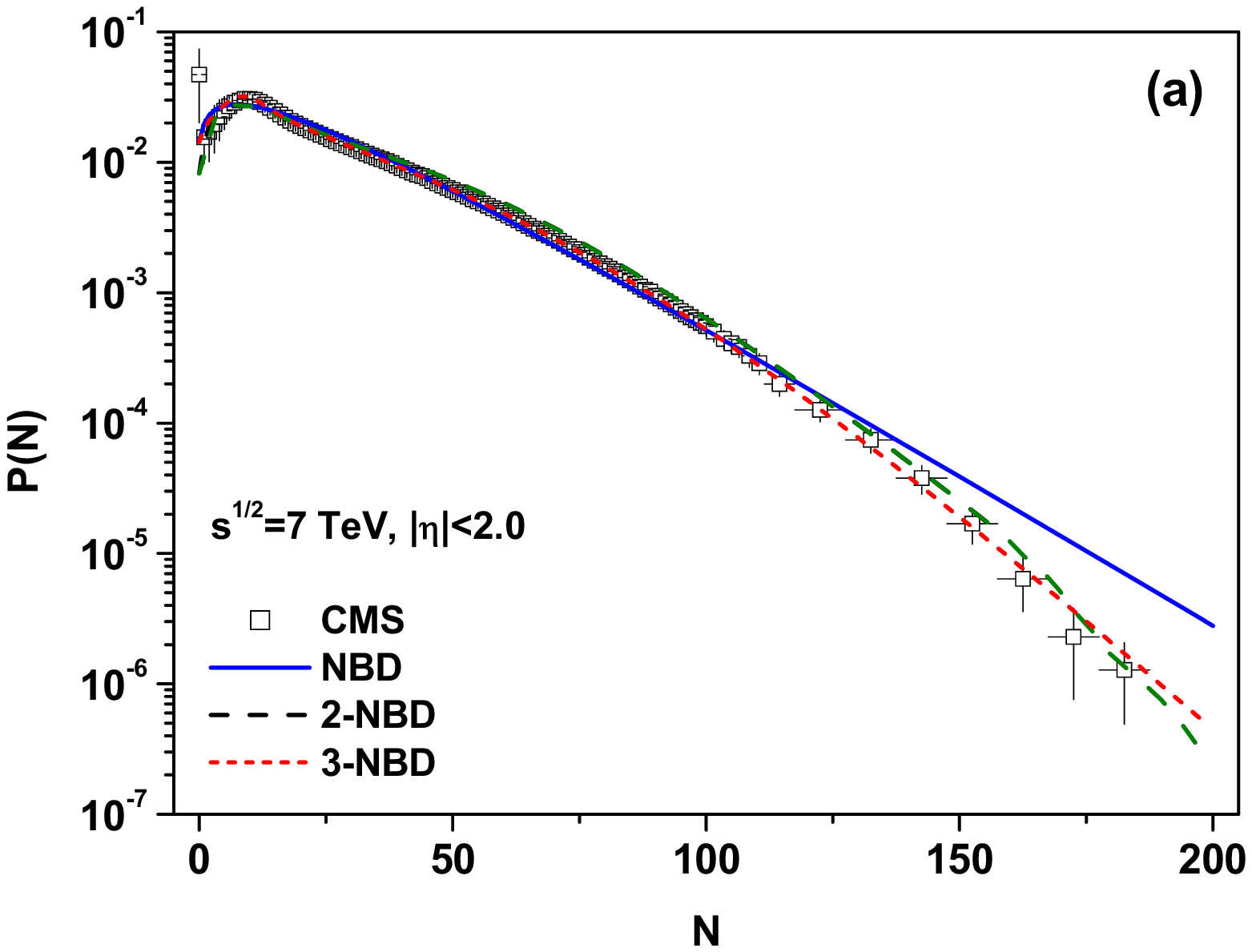}\hspace{10mm}
\includegraphics[scale=0.35]{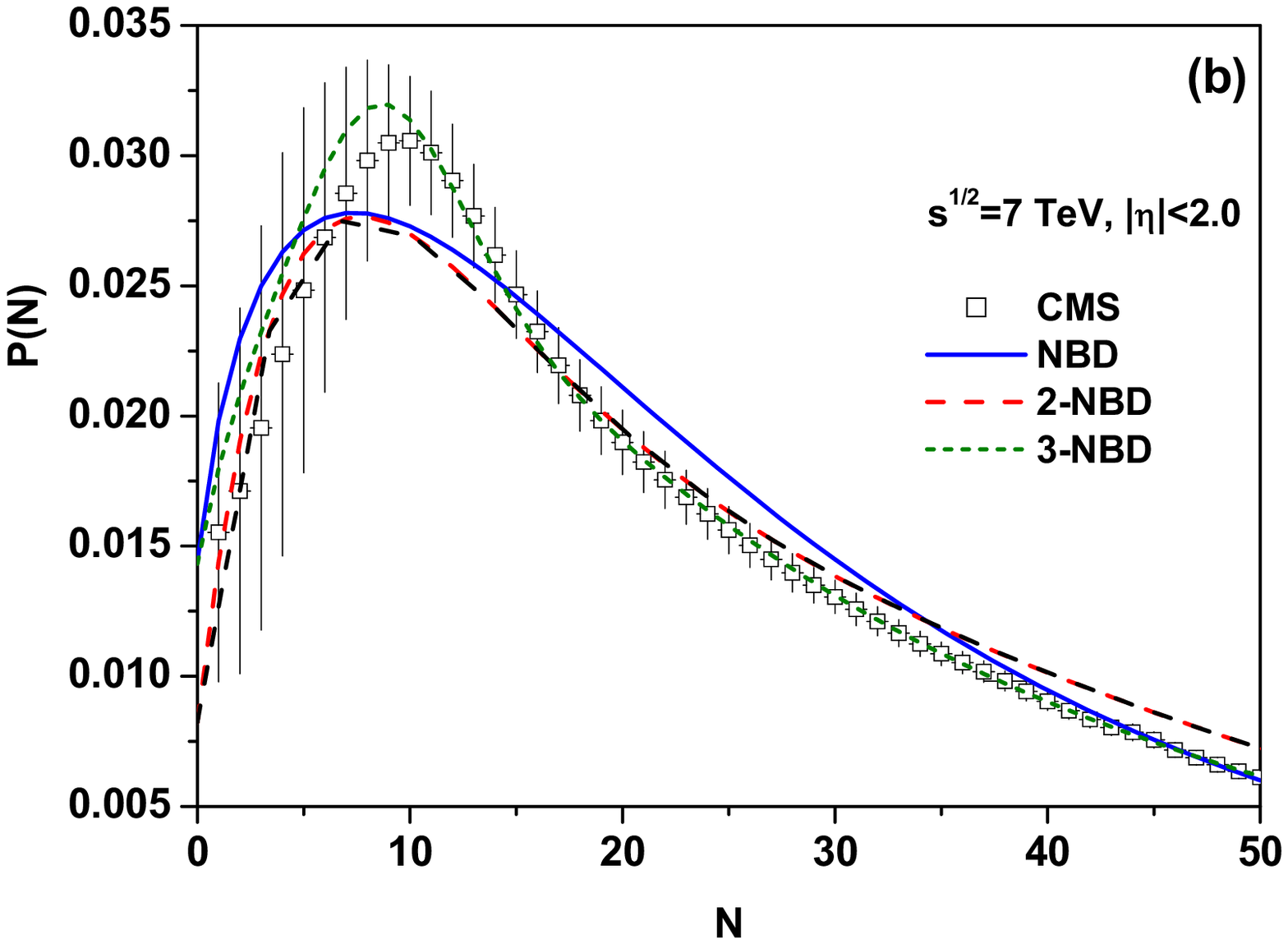}\\
\vspace{-5mm}
\includegraphics[scale=0.35]{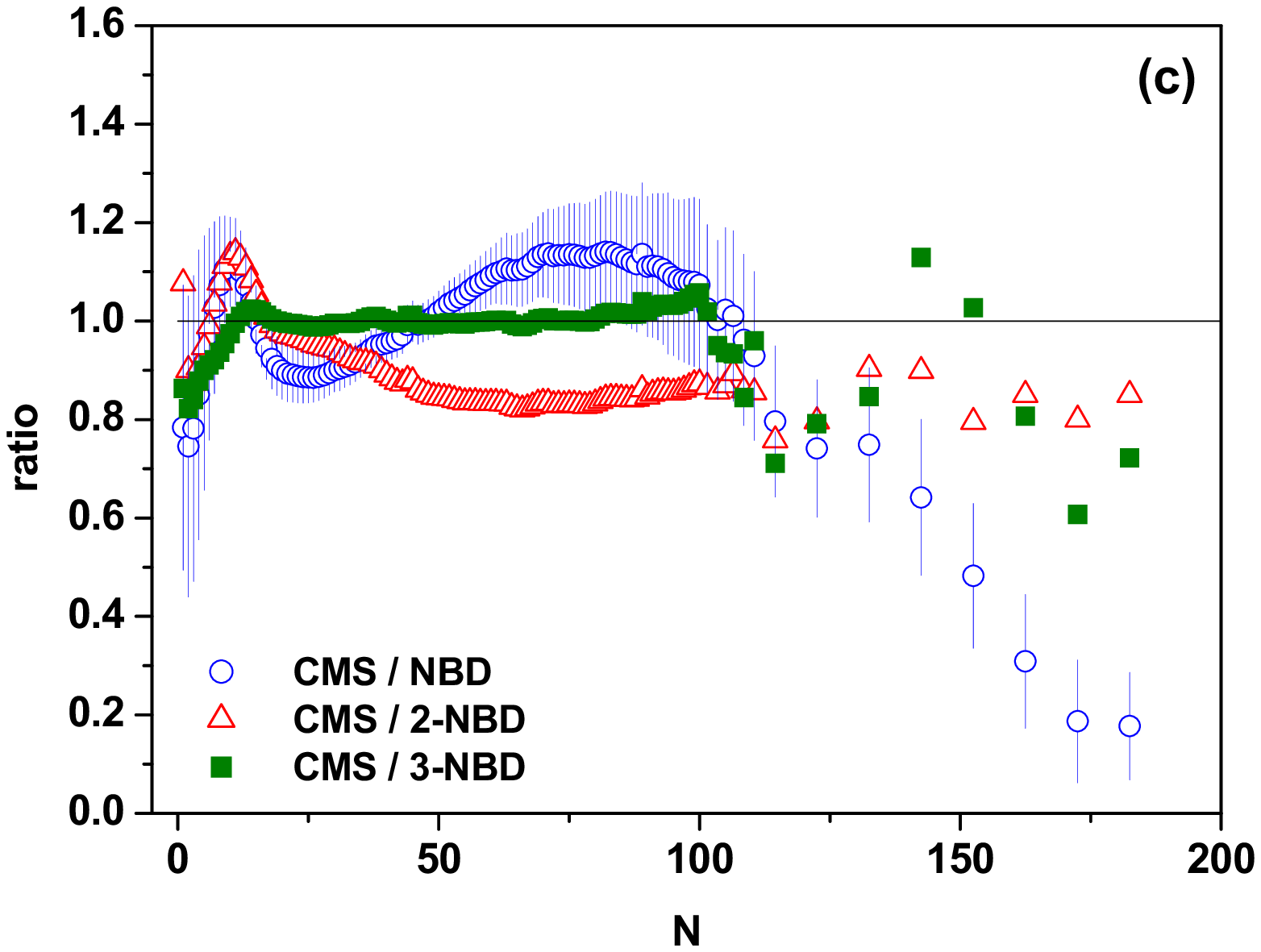}\hspace{10mm}
\includegraphics[scale=0.35]{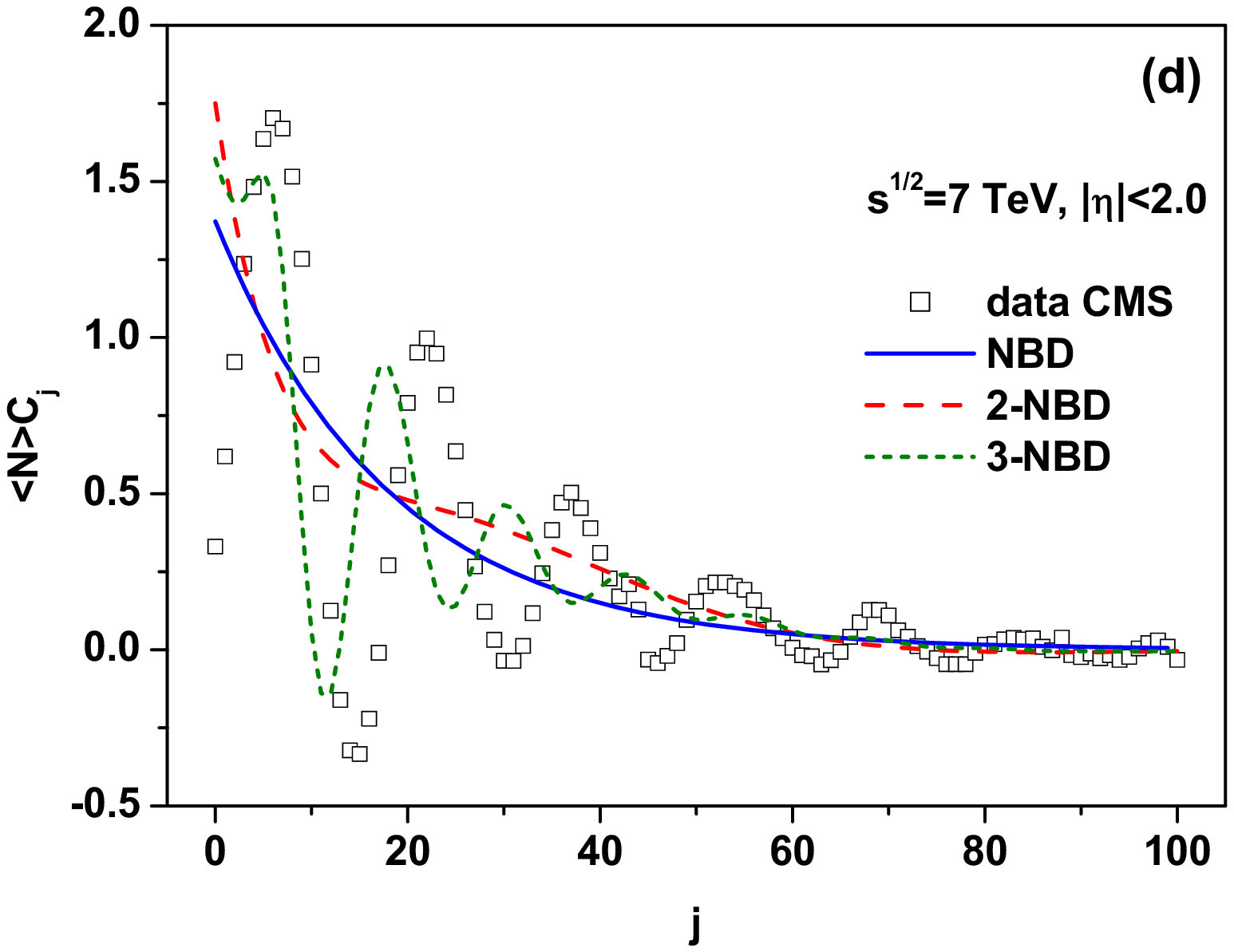}
\end{center}
\vspace{-10mm}
\caption{Results for $P(N)$, $(a)$; its enlarged part for $N < 50$, $(b)$; ratio $R=data/fit$, $(c)$, and for $C_j$, $(d)$, obtained  using a $3$-component $P(N)$ composed of $3$ NBD as proposed in \cite{Z} (with parameters the same as in \cite{Z}).}
\label{Zbor}
\end{figure}

\begin{figure}
\vspace{-7mm}
\begin{center}
\includegraphics[scale=0.35]{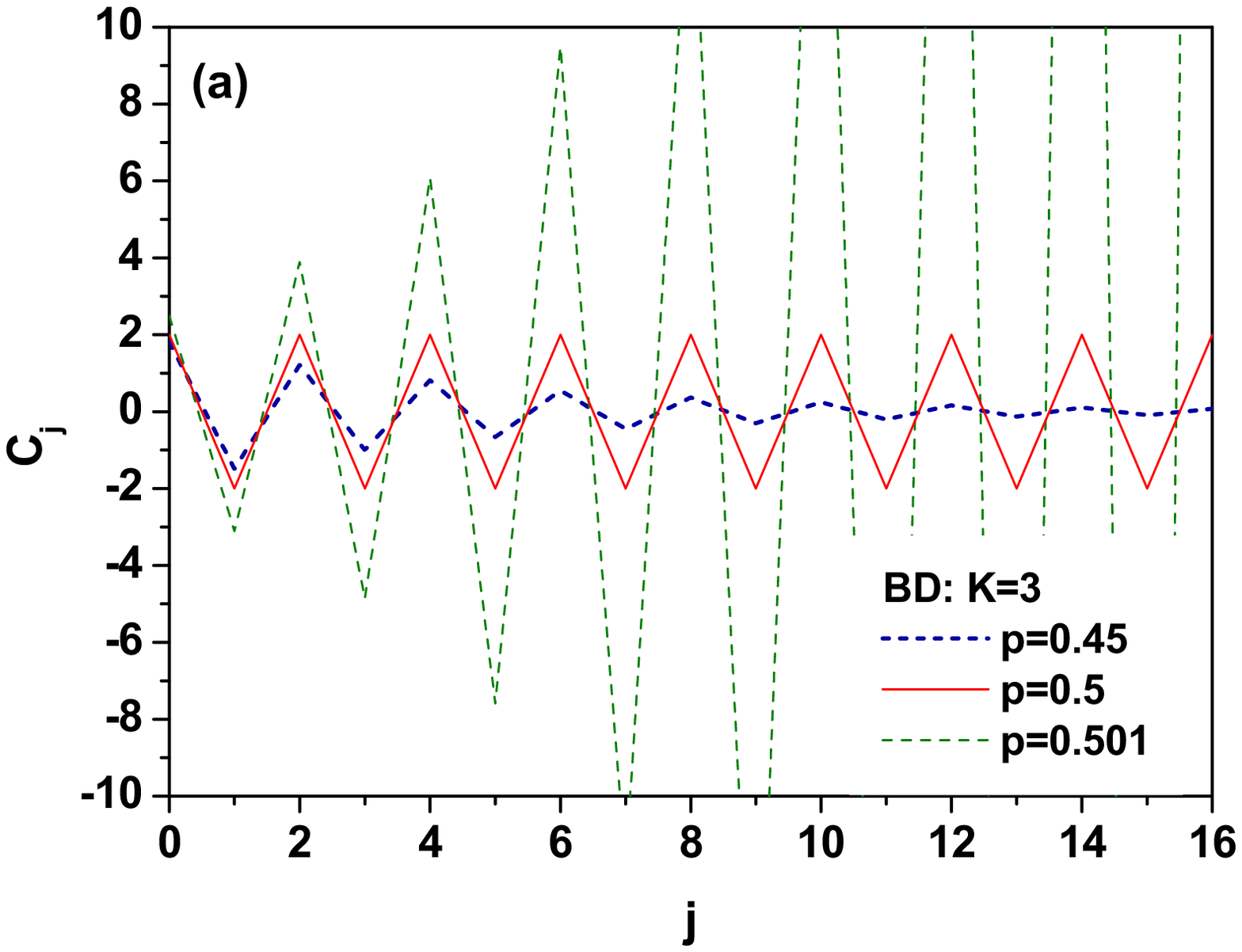}\hspace{10mm}
\includegraphics[scale=0.35]{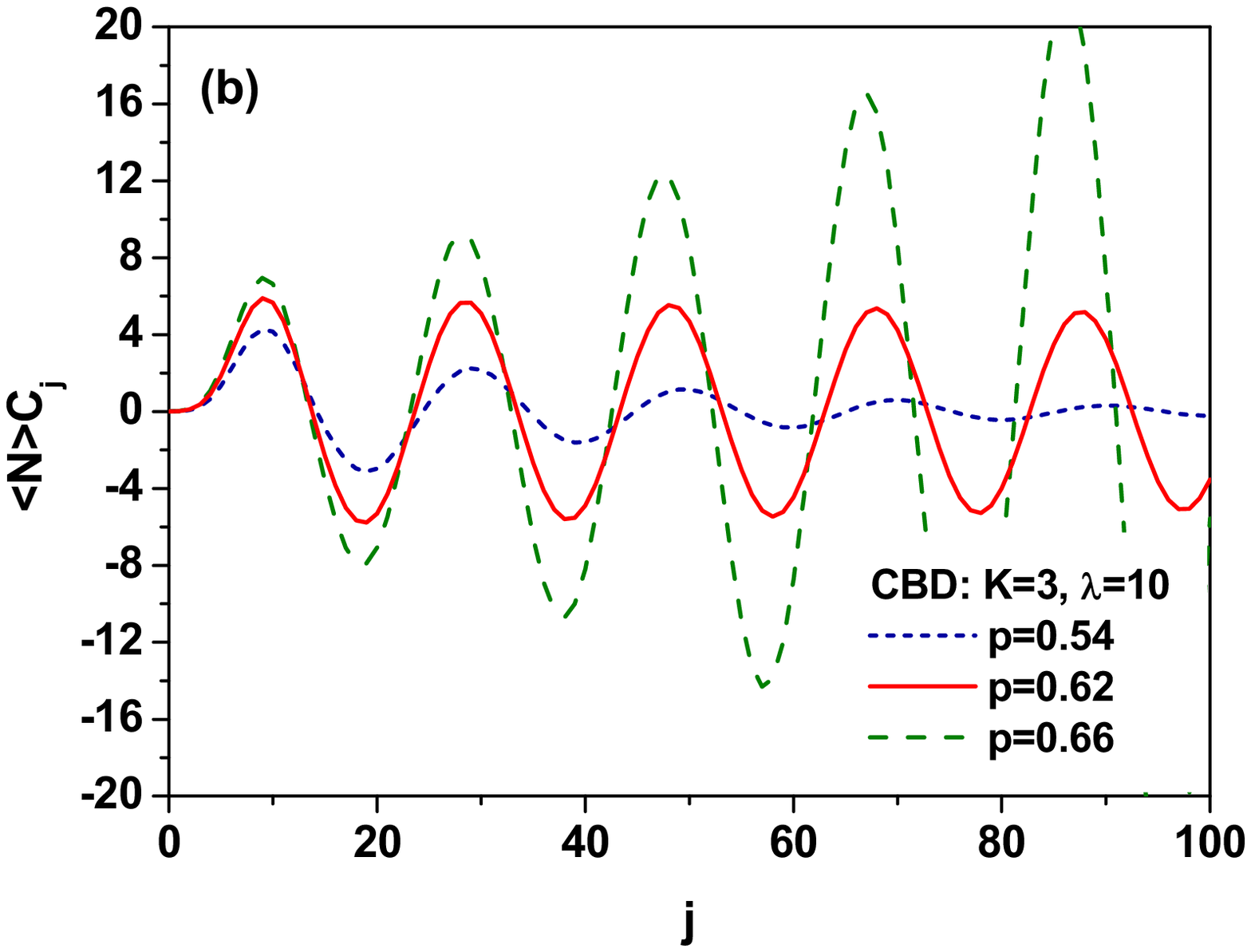}
\end{center}
\vspace{-10mm}
\caption{Examples of $C_j$ for: $(a)$ Binomial Distributions (BD) (Eq. (\ref{C_jBD})) and $(b)$ for the Compound Binomial Distributions (CBD) defined by Eq. (\ref{H_FG}).
}
\label{Fig3}
\vspace{-10mm}
\end{figure}

The above example shows that there is room for change in $P(N)$ resulting in agreement with data over the whole region of $N$. However, the recurrence relation (\ref{rr1}) is too restricted to be helpful in this case and in \cite{JPG} we proposed a more general form of the recurrence relation, that used in counting statistics when dealing with cascade stochastic processes \cite{ST}. Contrary to Eq. (\ref{rr1}), it now connects all multiplicities by means of some coefficients $C_j$ which define the corresponding $P(N)$ in the following way:
\begin{equation}
(N + 1)P(N + 1) = \langle N\rangle \sum^{N}_{j=0} C_j P(N - j). \label{Cj}
\end{equation}
The coefficients $C_j$ contain the memory of particle $N+1$ about all the $N-j$ previously produced particles. They can be directly calculated from the experimentally measured $P(N)$ by reversing Eq. (\ref{Cj}) and putting it in the form of the following recurrence formula \cite{JPG}:
\begin{equation}
\langle N\rangle C_j = (j+1)\left[ \frac{P(j+1)}{P(0)} \right] - \langle N\rangle \sum^{j-1}_{i=0}C_i \left[ \frac{P(j-i)}{P(0)} \right]. \label{rCj}
\end{equation}
The coefficients $C_j$ can therefore replace the ratio $R = data/fit$ in quality assessment of $P(N)$ used to fit data. The result was striking, as can be seen in Fig. \ref{Fig1} $(d)$, where the coefficients $C_j$ obtained from the data used in Fig. \ref{Fig1} $(a)$ show very distinct oscillatory behavior (with a period roughly equal to $16$), gradually disappearing with $N$. It turns out that it can be reproduced only by the MNB model \cite{JPG} mentioned before, which makes $R(N) =1$ for all $N$ \cite{JPG}.  As shown in \cite{JPG,ISMD16,IJMPA} such oscillations of $C_j$ are seen for different pseudorapidity windows, in data from all LHC experiments and energies. The only condition is that the statistics of the experiment must be high enough, for with small statistics the oscillations become too fuzzy to be recognized \cite{ISMD16,IJMPA} (the simplest way to obtain oscillatory behaviour of $C_j$ for an otherwise smooth distribution $P(N)$ is to distort $P(N)$ slightly at some point,  this distortion then propagates further \cite{JPG,IJMPA}). The reason why the single NBD is not able to reproduce data is that in this case all $C_j > 0$ \cite{JPG}:
\begin{equation}
C_j = \frac{k}{\langle N\rangle} p^{j+1} = \frac{k}{k + m} \exp( j \ln p). \label{CjNBD}
\end{equation}
The oscillations can occur only for combinations of NBD \cite{JPG,ISMD16}. However, the parameters of the $2$-NBD fit used in Fig. \ref{Fig1} $(a)$ result in a very small trace of oscillations of $C_j$ and we were not able to find parameters of a $2$-NBD allowing for a reasonable description of $P(N)$ and $C_j$ at the same time. The best result obtained so far is the $3$-NBD fit proposed in \cite{Z} and based on the claim that there is a place in data for a third component aiming to describe the low $N$ events (see \cite{Z} for details, it agrees with our observations mentioned before). Fig. \ref{Zbor} shows the results obtained for the parameters from \cite{Z}. Note that the agreement of $P(N)$ with data and the behavior of both the ratio $R = data/fit $ and the coefficients $C_j$ improved substantially. However, as one can see in Fig. \ref{Zbor} $(b)$, the low $N$ region of $P(N)$ still shows some deviations, which, albeit rather small, result in $R$ departing from unity (downwards) at small $N$ and in $C_j$ missing the data for large $j$.
Contrary to the case of the NBD, the modified combinants for the BD, cf. Eq. (\ref{BD}), oscillate rapidly,
\begin{equation}
C_j = (-1)^j \frac{K}{\langle N\rangle} \left( \frac{\langle N\rangle}{K-\langle N\rangle}\right)^{(j+1)} = \frac{(-1)^j}{1 - p}\left( \frac{p}{1 - p}\right)^{j}, \label{C_jBD}
\end{equation}
with a period equal to $2$. In Fig. \ref{Fig3} $(a)$ one can see that the amplitude of these oscillations depends on the emission probability $p$, in this case the $C_j$ increase with rank $j$ for $p > 0.5$ and decrease for $p < 0.5$ (this is, however, not generally true as we shall see later, cf., for example Fig. \ref{Fig3} $(b)$). However, their general shape lacks the distinctively fading down feature of the $C_j$ observed experimentally. This means that the BD used alone cannot explain the data (see also \cite{GMD})\footnote{Some comments concerning the credibility of using $C_j$ and on their oscillations are necessary at this point. In a recent review \cite{Alkin} the observed oscillations were attributed to the possible peculiarities of the experimental unfolding procedure used while preparing the final data. However, such a statement has so far not been substantiated by any known experimental analysis of the procedure used, furthermore, the peculiarities seen in the ratio $R$ (tightly connected with the oscillations of $C_j$) were not addressed as well. Therefore we assumed that this is a real new effect, connected with some dynamical features of the production mechanism (in fact in \cite{CSF,CSF1} the cascade stochastic processes leading to Eq. (\ref{rCj}) were successfully applied to multiparticle phenomenology).}.

It turns out that the coefficients $C_j$ defined by {\it the recurrence relation} (\ref{rCj}) are closely related to the so called {\it combinants} $C^{\star}_j$ which are defined in terms of {\it the generating function}, $G(z)=\sum^{\infty}_{N=0} P(N) z^N $, as
\begin{equation}
C^{\star}_j = \frac{1}{j!} \frac{d^j \ln G(z)}{d z^j}\bigg|_{z=0}\qquad {\rm or}\qquad \ln G(z) = \ln P(0) + \sum^{\infty}_{j=1} C^{\star}_j z^j, \label{CombDef}
\end{equation}
and which were introduced in \cite{KG} (see also \cite{Kittel,CombUse,H309,H318,H463,Astro,Book-BP}). Namely,
\begin{equation}
C_j = \frac{j+1}{\langle N\rangle} C^{\star}_{j+1}. \label{connection}
\end{equation}
Therefore, henceforth we shall call the $C_j$ {\it modified combinants}.  Note that the recurrence relation, Eq. (\ref{Cj}) can be  written in terms of $C_j^{\star}$ as:
\begin{equation}
(N + 1)P(N + 1) = \sum^{N}_{j=0} (j+1)C_{j+1}^{\star} P(N - j), \label{Cj_star}
\end{equation}
and that the $C_j$ can be expressed by the generating function $G(z)$ of $P(N)$ as
\begin{equation}
\langle N\rangle C_j = \frac{1}{j!} \frac{ d^{j+1} \ln G(z)}{d z^{j+1}}\bigg|_{z=0}. \label{GF_Cj}
\end{equation}
This relation will be used in what follows when calculating $C_j$ from multiplicity distributions $P(N)$ defined by some generating function $G(z)$.

Because a single distribution of the NBD or BD type cannot describe data we shall check the idea of {\it compound distributions} (CD) applicable when the production process consists of a number $M$ of some objects (clusters/fireballs/etc.) produced according to some distribution $f(M)$ (defined by a generating function $F(z)$), which subsequently decay independently into a number of secondaries, $n_{i = 1,\dots, M}$, following some other (always the same for all $M$) distribution, $g(n)$ (defined by a generating function $G(z)$) \cite{Compound}\footnote{In fact the NBD is a compound Poisson distribution with the number of clusters given by a Poissonian distribution and the particles inside the clusters distributed according to a logarithmic distribution \cite{GVH}. In \cite{BSWW} we proposed a specific compound distribution to explain the Bose-Einstein correlation phenomenon. It consisted of a  combination of $k$ {\it elementary emitting cells (EEC)} producing particles according to a geometrical distribution. For $k=const$  the resultant $P(N)$ was of the NBD type, for $k$ distributed according to a BD it was a modified NBD. However, applying it to the present situation we could not find a set of parameters providing both the observed $P(N)$ and oscillating $C_j$.}. The resultant multiplicity distribution, $h( N ) = f\otimes  g$, where $ N =\sum_{i=0}^M n_i$, is a compound distribution of $f$ and $g$ with generating function\footnote{Note that for the class of distributions of $M$ that satisfy the recurrence  relation Eq. (\ref{rr1}) the compound distribution $h=f\otimes  g$ is also given by the so-called Panjer's recurrence relation \cite{Panjer}, $ Nh(N) = \sum^{N}_{j=1}[ \beta N + (\alpha - \beta)j ] g(j) h(N-j) = \sum^{N}_{j=1} C^{(P)}_j (N) h(N-j)$, with initial value $h(0)=f(0)$. It could be considered as a generalization of Eq. (\ref{Cj}) used to define modified combinants with coefficients $C^{(P)}_j$ depending additionally on $N$. However, Eq. (\ref{Cj}) is not limited to the class of distributions satisfying Eq.(\ref{rr1}) but is valid for any distribution $P(N)$, therefore this recursion relation is not suitable for us.}.
\begin{equation}
H(z) = F[G(z)] \label{CD_GF}
\end{equation}
for which
\begin{equation}
\langle N\rangle = \langle M\rangle \langle n\rangle\qquad {\rm and}\qquad Var(N) = \langle M\rangle Var(n) + Var(M) \langle n\rangle^2 .\label{CD_moments}
\end{equation}
Let us take, as an example, $f$ as a Binomial Distribution with generating function $F(z) = (pz + 1 - p)^K$ (for which the $C_j$ oscillate with a period of $2$, cf. Fig. \ref{Fig3}$(a)$), and  $g$ as a Poisson distribution with generating function $G(z) = \exp[ \lambda (z-1)]$ (for which $C_0 = 2$ and $C_{j>0} = 0$, cf. Eq. (\ref{GF_Cj})). The generating function of the resulting Compound Binomial Distribution (CBD) is then
\begin{equation}
H(z) = \left\{ p \exp[ \lambda (z-1)] + 1 -p \right\}^K. \label{H_FG}
\end{equation}
The analytical forms of the corresponding  $C_j$ and $P(N)$ are presented in \cite{IJMPA} (as Eqs. (136)-(138)). Fig. \ref{Fig3} $(b)$ shows $C_j$ for the CBD with $K=3$ and $\lambda =10$ calculated for three different values of $p$ in the BD: $p=0.54,~0.62,~0.66$. Note that, in general, the period of the oscillations is now equal to $2\lambda$ (i.e., in Fig. \ref{Fig3} $(b)$ where $\lambda = 10$ it is equal to $20$).  This example shows that the choice of a BD as the basis of the CD used is crucial to obtain the oscillatory behavior of the $C_j$ (for example, a compound distribution formed from a NBD and some other NBD provides smooth $C_j$).
\begin{figure}[t]
\begin{center}
\includegraphics[scale=0.35]{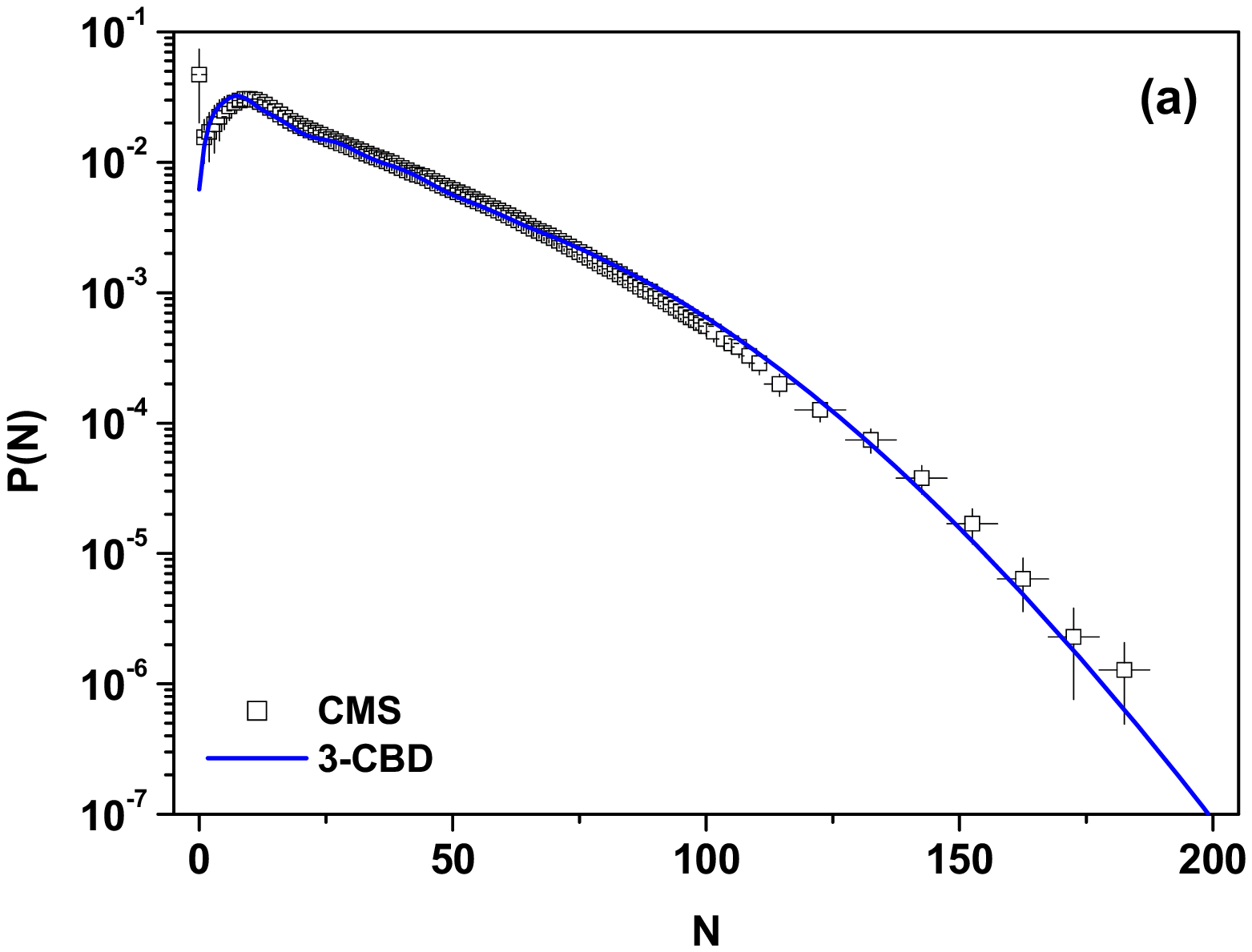}\hspace{10mm}
\includegraphics[scale=0.35]{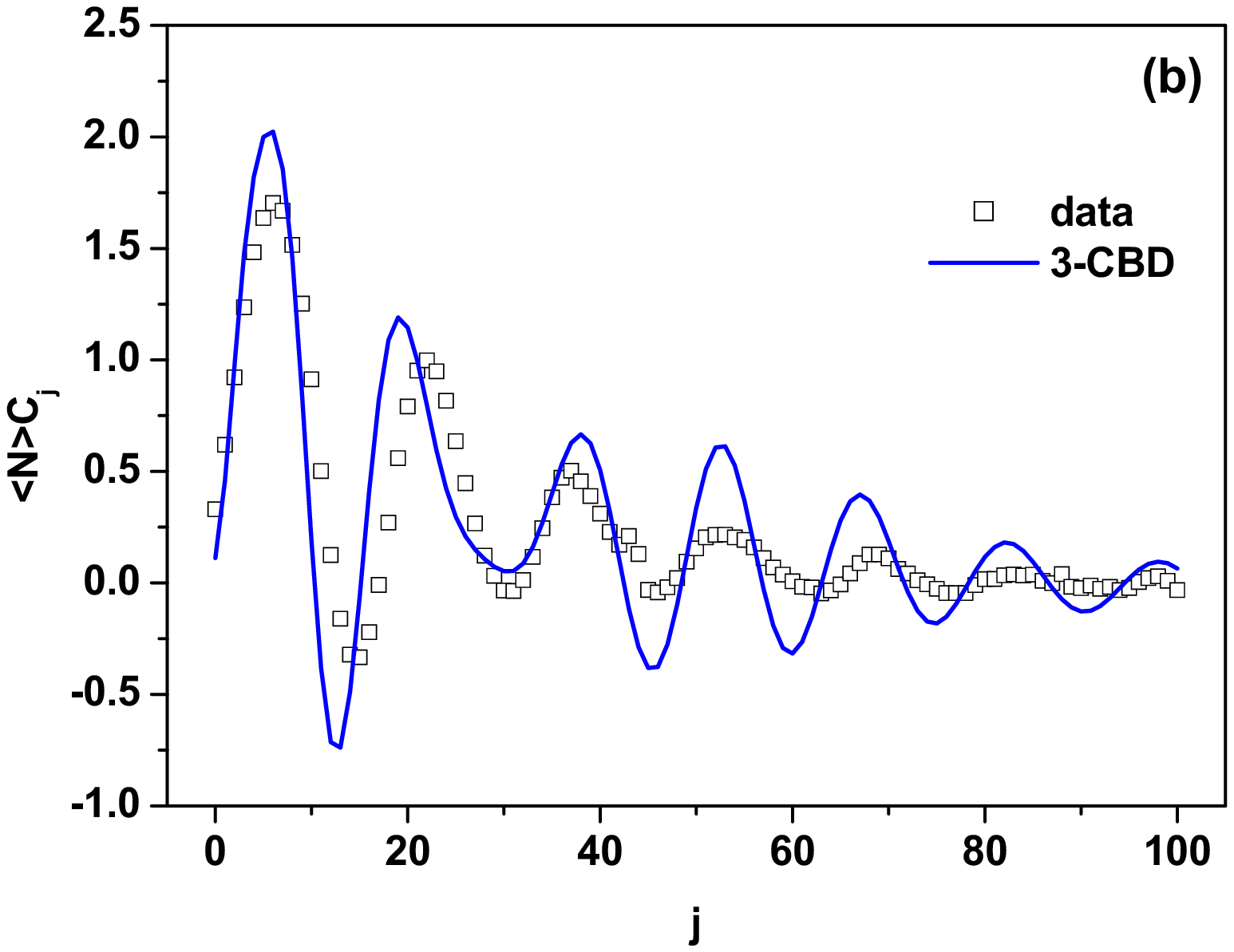}\\
\vspace{-5mm}
\includegraphics[scale=0.35]{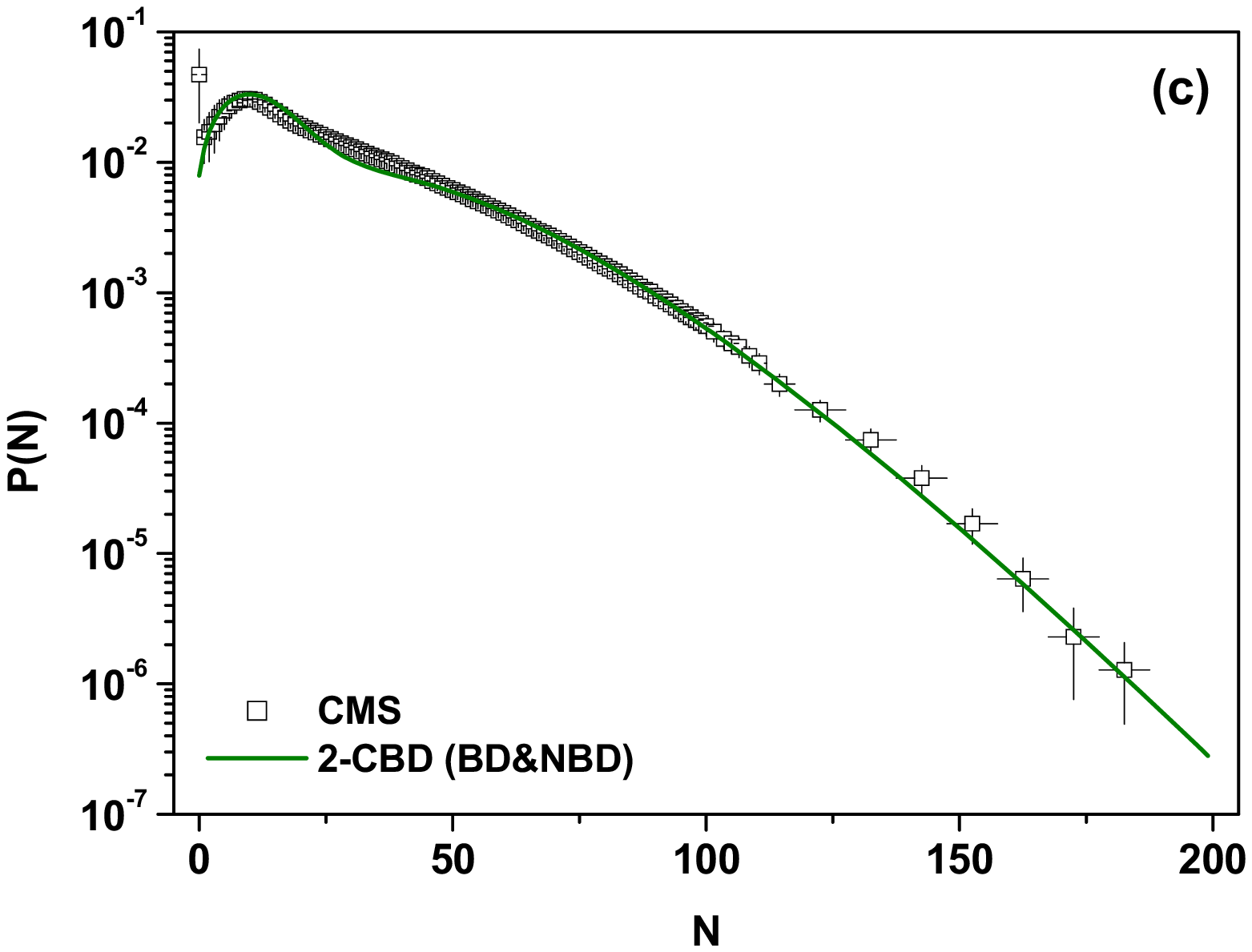}\hspace{10mm}
\includegraphics[scale=0.35]{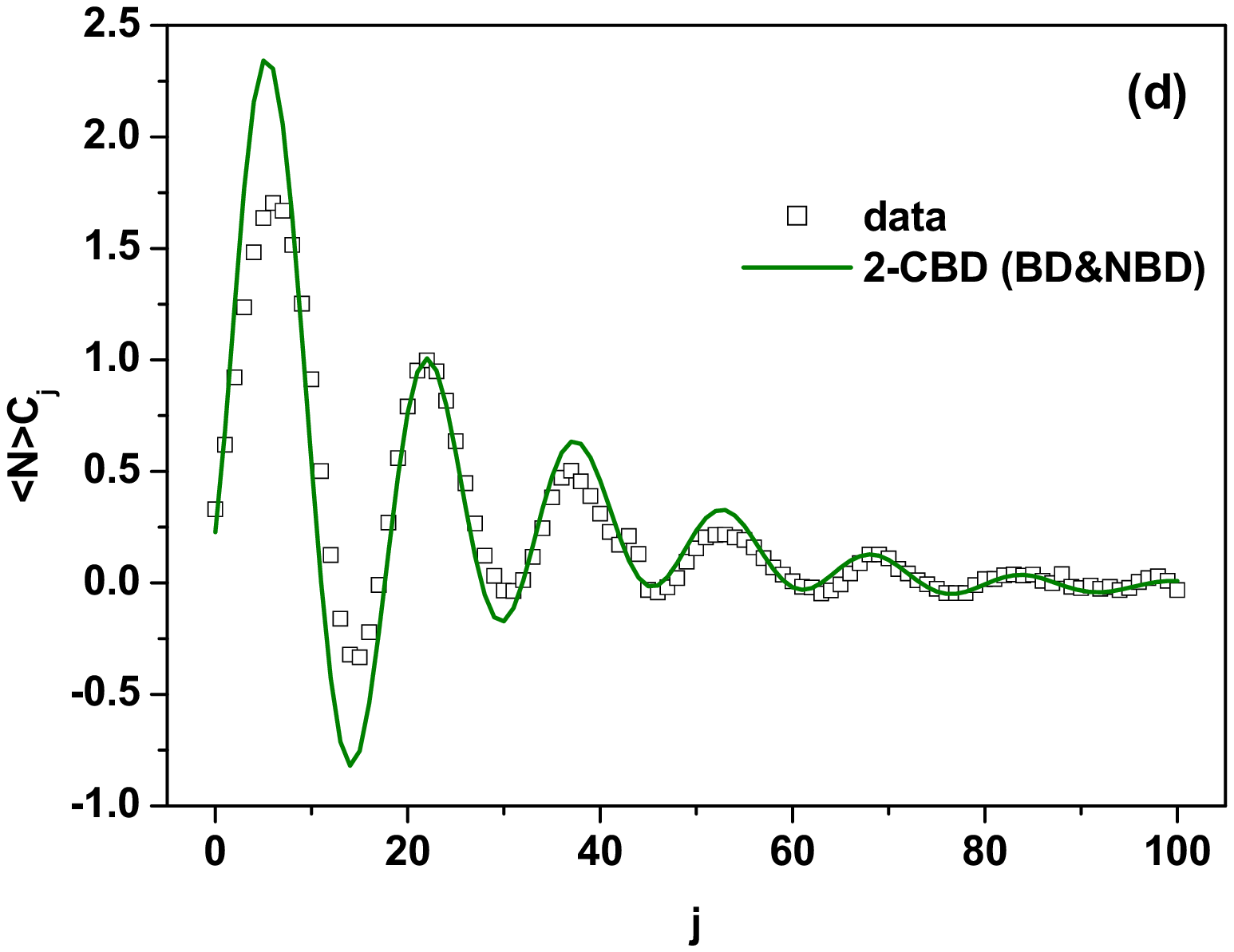}
\end{center}
\vspace{-10mm}
\caption{$(a)$  Charged hadron multiplicity distributions for $|\eta| < 2$ at $\sqrt{s} = 7$ TeV, as given by the CMS experiment \cite{CMS} (points), compared with a $3$-component CBD, Eq. (\ref{3CBD}). $(b)$ Coefficients $C_j$ emerging from the CMS data used in panel $(a)$ compared with the corresponding $C_j$ obtained from the  $3$-component compound binomial distribution ($3$-CBD). $(c)$  Charged hadron multiplicity distributions for $|\eta| < 2$ at $\sqrt{s} = 7$ TeV, as given by the CMS experiment \cite{CMS} (points), compared with a $2$-component CBD, Eq. (\ref{2CBD}). $(d)$ Coefficients $C_j$ emerging from the CMS data used in panel $(c)$ compared with the corresponding $C_j$ obtained from the  $2$-component compound binomial distribution ($2$-CBD).
}
\label{Fig4}
\vspace{-5mm}
\end{figure}

Unfortunately, such a single component CBD (with $P(N)=h(N;p,K,\lambda)$ depending on three parameters: $p$, $K$ and $\lambda)$, does not describe the experimental $P(N)$ and $C_j$. We return therefore to the idea of using a multicomponent version of the CBD presenting two examples. The first is $3$-component CBD defined as:
\begin{equation}
P(N) = \sum_{i=1,2,3} w_i h\left(N; p_i, K_i, \lambda_i\right);\qquad \qquad \sum_{i=1,2,3} w_i = 1.  \label{3CBD}
\end{equation}
The results of using Eq. (\ref{3CBD}) (with parameters: $\omega_1 = 0.34$, $\omega_2 = 0.4$, $\omega_3 = 0.26$; $p_1 = 0.22$, $p_2 = 0.22$, $p_3 = 0.12$; $K_1 = 10$, $K_2 = 12$, $K_3 = 30$ and $\lambda_1 = 4$, $\lambda_2 = 9$, $\lambda_3 = 14$) are presented in Figs. \ref{Fig4} $(a)$ and $(b)$. As one can see, this time the fit to the $P(N)$ is quite good and the modified combinants $C_j$ follow the oscillatory pattern as far as the period of the oscillations is concerned, albeit their amplitudes still decay too slowly. To improve this deficiency we present, as a second example, a $2$-component version of the CBD in which the Poisson distribution has been replaced by a NBD. Its generating function is
\begin{equation}
H(z) = \left[ p\left( \frac{1 - p'}{1 - p'z}\right)^k + 1 - p\right]^K,\qquad {\rm where}\qquad p' = \frac{m}{m + k}, \label{2-CBD}
\end{equation}
and
\begin{equation}
P(N) = \sum_{i=1,2} w_i h\left(N; p_i, K_i, k_i, m_i\right);\qquad \qquad \sum_{1=1,2} w_i = 1.  \label{2CBD}
\end{equation}
As one can see in Figs. \ref{Fig4} $(c)$ and $(d)$, using Eq. (\ref{2CBD}) (with parameters:  $K_1 = K_2 = 3$, $p_1 = 0.7$, $p_2 = 0.67$, $k_1 = 4$, $k_2 = 2.3$, $m_1 = 6$, $m_2 = 19.0$ and $w_1 = w_2 = 0.5$) improves substantially the behavior of $C_j$. This means that to describe data one has to use some multicomponent compound distributions based on the BD (as responsible for the oscillations in $C_j$) and some other distribution providing damping of the oscillations for large $N$ (here the NBD).

\begin{figure}[h]
\vspace{-3mm}
\begin{center}
\includegraphics[scale=0.35]{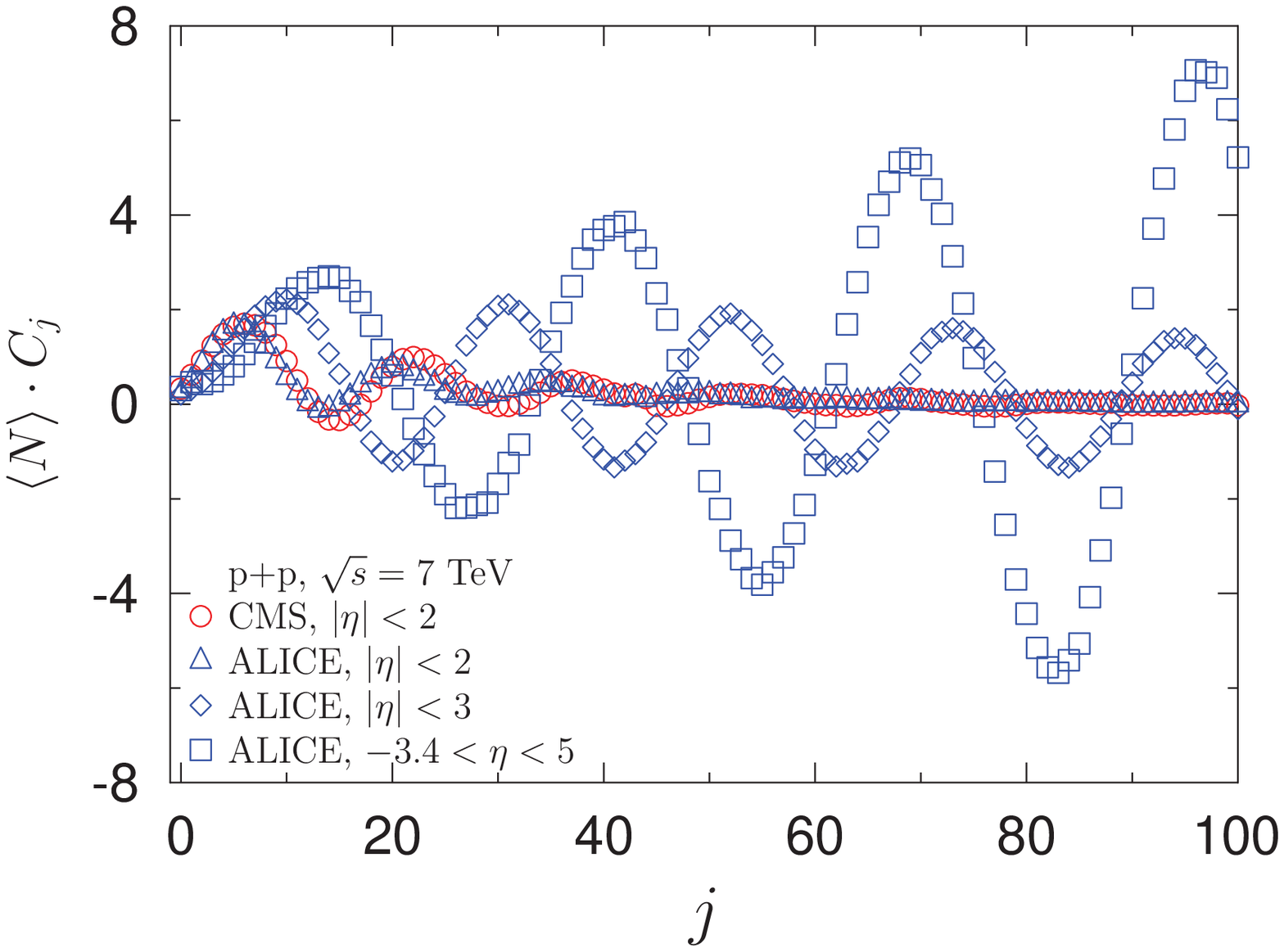}\hspace{10mm}
\includegraphics[scale=0.35]{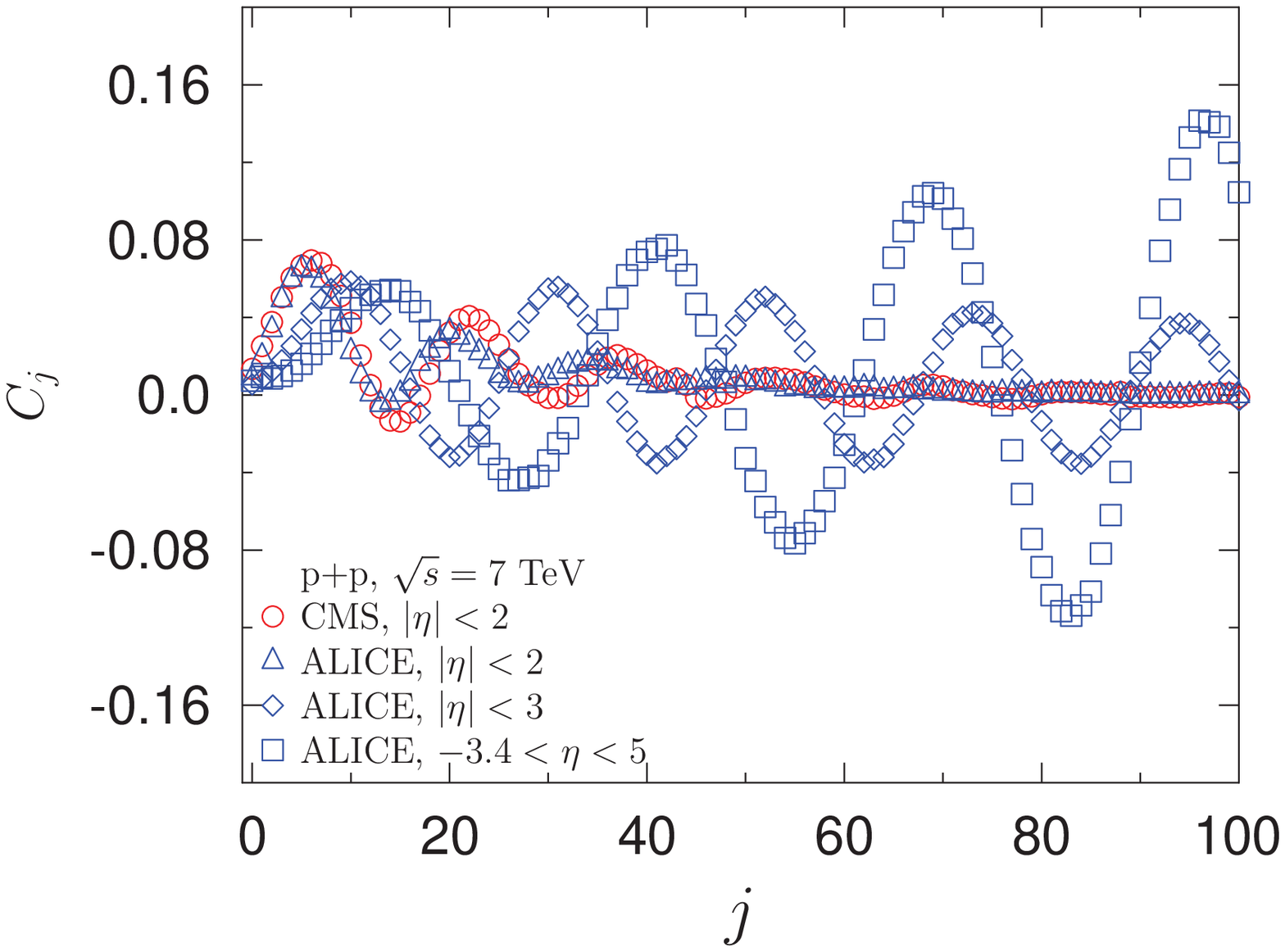}
\end{center}
\vspace{-5mm}
\caption{Comparison of $\langle N\rangle C_j$ and $C_j$ obtained from the $P(N)$ measured by CMS \cite{CMS} and ALICE \cite{ALICE} for three different rapidity windows. See text for details.}
\label{XXX}
\vspace{-5mm}
\end{figure}
With such knowledge we proceeded to a description of the recent ALICE data \cite{ALICE} for multiplicity distributions (NSD events at $7$ TeV), at three different rapidity windows: $|\eta| < 2$, $|\eta|<3$ and $-3.4<\eta<5$. In Fig. \ref{XXX} we show on the left panel the results for $\langle N\rangle C_j$ obtained from the measured $P(N)$; for comparison the previously used CMS data \cite{CMS} for $|\eta|<2$ are also shown (and they agree with the data from ALICE). The most intriguing features observed is the rather dramatic increase of both the period of the oscillations and their amplitude with the width of the rapidity window used to collect the data and, most noticeably, the previously observed fading down of their amplitude is now replaced by an (almost) constant behavior (for $|\eta| < 3$) and by a rather dramatic increase (for $-3.4 < \eta < 5$). Because, roughly, $\langle N\rangle \sim \Delta \eta $, one would expect that at least part of the increase of amplitude could come from the increase of $\langle N\rangle$ with $\Delta \eta$. The right panel of Fig. \ref{XXX} shows that this can only partially be true, the previously observed effects remain, albeit they are perhaps not so dramatic as before.

\begin{figure}[t]
\begin{center}
\includegraphics[scale=0.65]{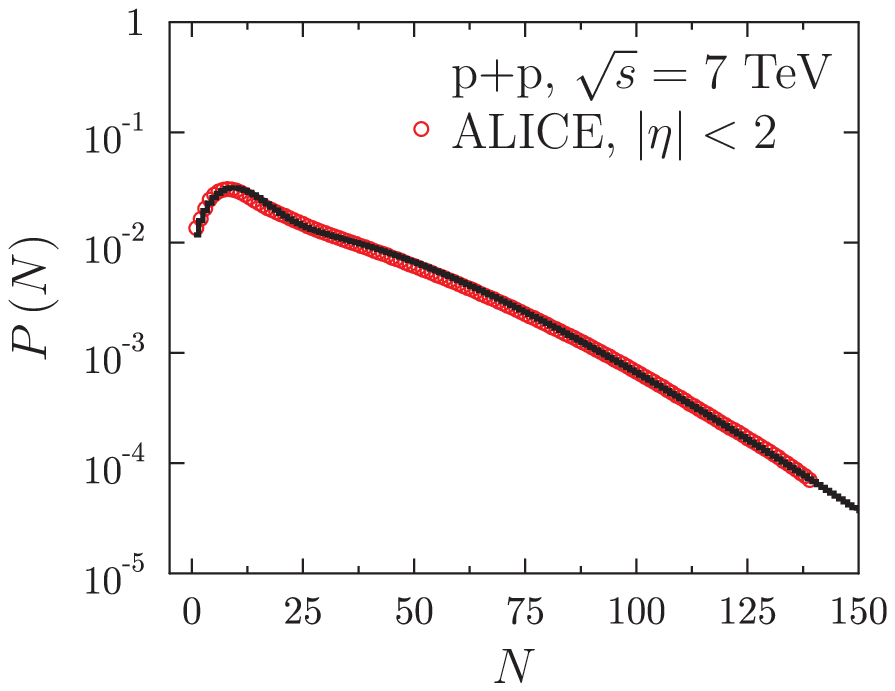}\hspace{5mm}
\includegraphics[scale=0.65]{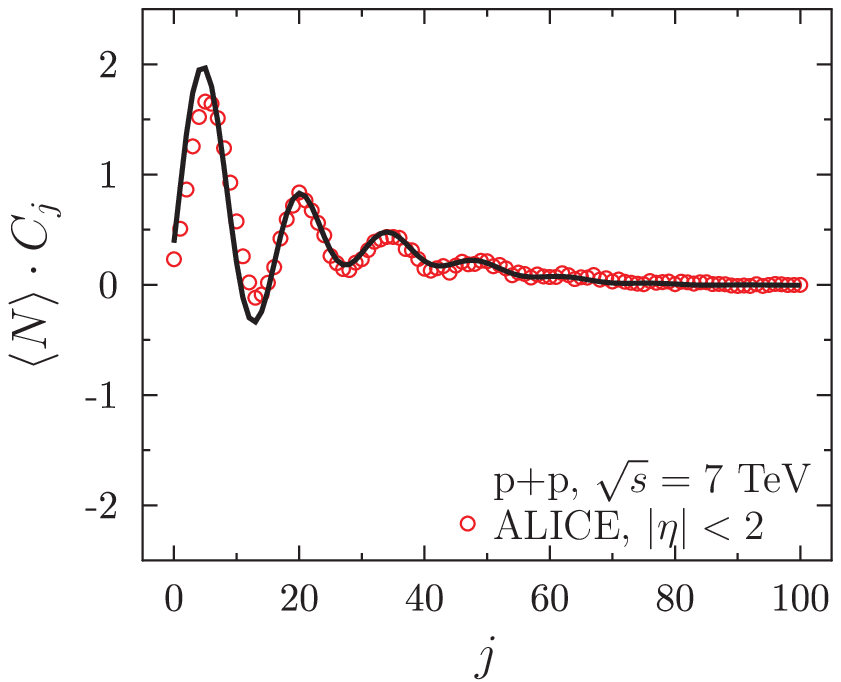}\\
\includegraphics[scale=0.65]{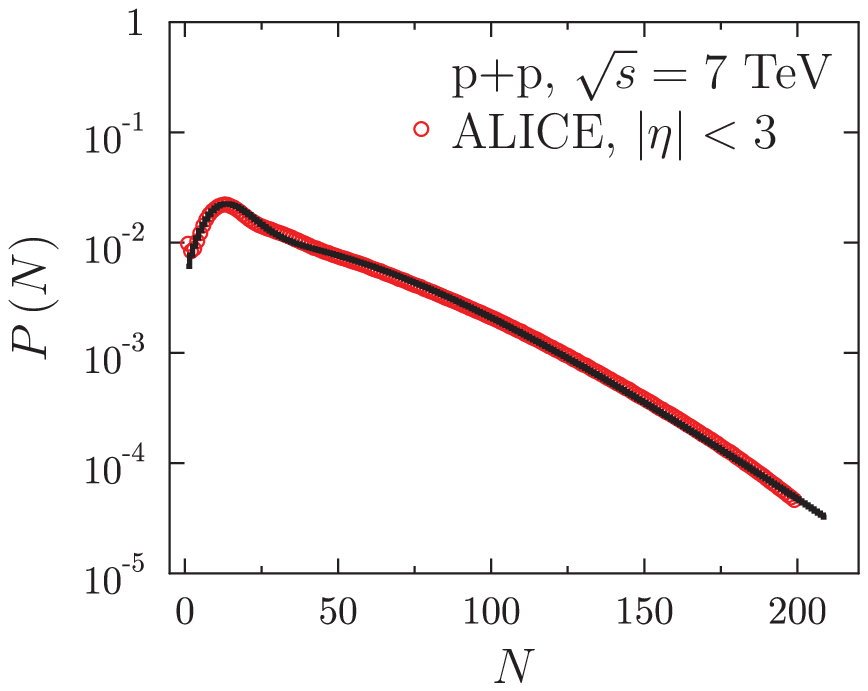}\hspace{5mm}
\includegraphics[scale=0.65]{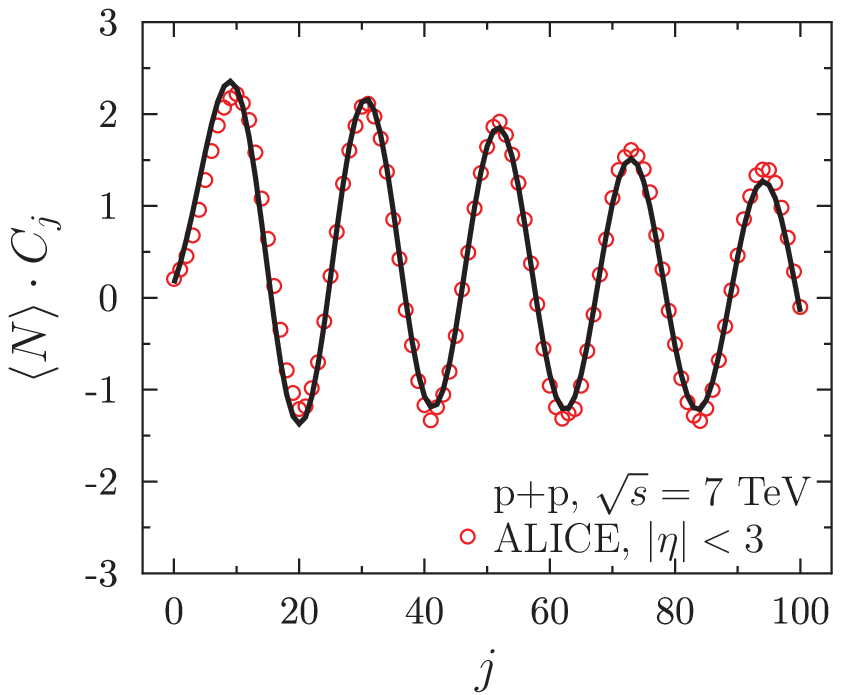}\\
\includegraphics[scale=0.65]{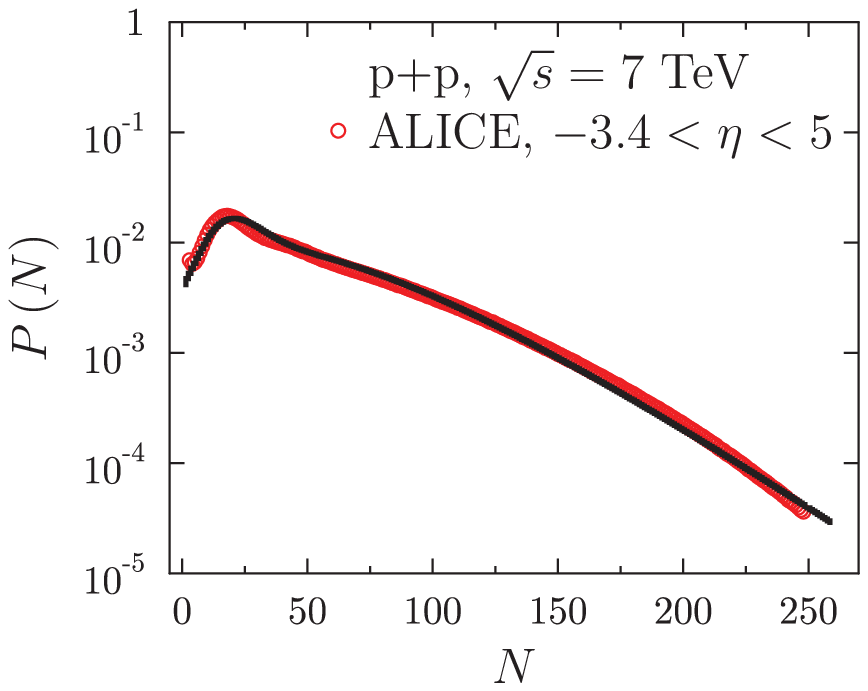}\hspace{5mm}
\includegraphics[scale=0.65]{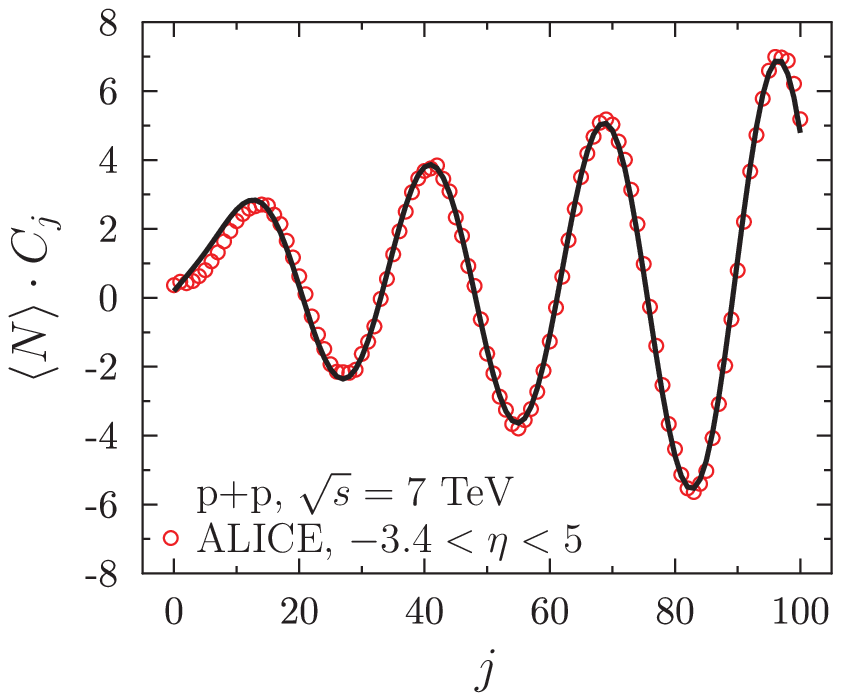}
\end{center}
\vspace{-5mm}
\caption{Multiplicity distributions $P(N)$ (left panels) measured by ALICE \cite{ALICE} and modified combinants $C_j$ emerging from them (right panels) for three different rapidity windows fitted using the two compound distribution (BD+NBD) given by Eqs. (\ref{2-CBD}) and (\ref{2CBD}) with parameters listed in Table \ref{tab-1} (see text for details).}
\label{Fig5}
\vspace{-5mm}
\end{figure}
\begin{table}[b]
\centering
\caption{Parameters $w_i$, $p_i$, $K_i$, $k_i$ and $m_i$ of the $2$-component $P(N)$, Eqs. (\ref{2-CBD}) and (\ref{2CBD}), used to fit the data in Fig. \ref{Fig5}. For completeness $p'_i = m_i/(m_1 + k_i)$ from Eq. (\ref{2-CBD}) are also included.}
\label{tab-1}
\begin{tabular}{|c|cccccc|cccccc|}
\hline
                                   & $w_1$      & $p_1$  & $K_1$  & $k_1$ & $m_1$  & $p'_1$ & $w_2$  & $p_2$  &$K_2$& $k_2$ & $m_2$ & $p'_2$  \\\hline
 $-2\langle~\eta~\langle 2$        & $0.30$    & $0.75$  & $3$    & $3.8$ & $4.75$ & $0.56$ & $0.70$ & $0.70$ & $3$ & $1.30$ & $15.9$ & $0.924$ \\
 $-3\langle~\eta~\langle 3$        & $0.24$    & $0.90$  & $3$    & $2.8$ & $5.75$ & $0.67$ & $0.76$ & $0.645$ & $3$ & $1.34$ & $23.5$ & $0.946$ \\
 $-3.4\langle~\eta~\langle 5$      & $0.20$    & $0.965$  & $3$    & $2.7$ & $8.00$ & $0.75$ & $0.80$ & $0.72$ & $3$ & $1.18$ &
 $27.0$ & $0.955$ \\\hline
\end{tabular}
\end{table}

We close our presentation by showing in Fig. \ref{Fig5} that the results presented in Fig. \ref{XXX} can be nicely fitted, including (not shown there) $P(N)$ from \cite{ALICE}, using the two component CBD (BD+NBD) discussed above (Eqs. (\ref{2-CBD}) and (\ref{2CBD}) with the parameters listed in Table \ref{tab-1}. When performing these fits we were only concerned to find such values of the parameters as would reproduce the results in the best possible way (although no $\chi^2$ estimations were used). This means that, at the moment, the values presented in Table \ref{tab-1} must be taken {\it cum grano salis}. In general one observes a slow increase of $p$ and $p'$ (but with some nonmonotonicity seen for $p_2$), already observed in Fig. \ref{Fig3}. A more detailed analysis would need much more involved investigations than presented here at the moment and is currently under investigation. At the moment we must admit that the problem of the physical meaning of these fits (in other words: what information on the mechanism of production of particles they convey) remains still an open one.

\section{Some explanatory remarks}
\label{Summary}

Note that in all the above discussions we always remained on the same phenomenological level, either modifying the emission probability $p$ in the NBD, or combining it with some other distribution. So far we have not looked for physical justifications of the methods used but concentrated on reproducing $P(N)$ and the ratio $R=data/fit$ or the modified combinants $C_j$. The only more theoretically oriented attempt is Ref. \cite{BS} describing the pion multiplicity using the combinants $C^{\star}_j$ and discussing two scenarios. The first one had $N$ sources emitting bosons without any restrictions on their number, and resulted in a NBD and smooth and diminishing combinants. The second one had $M$ sources, each emitting only a limited number of bosons (one or two in \cite{BS}), and this resulted in a BD and oscillating combinants. We expect therefore that our $C_j$ would follow the same behavior. Note also that the NBD belongs to the class of the so-called infinite divisible distributions whereas the BD does not. In \cite{Book-BP} it is claimed that combinants of all ranks are all non-negative if and only if the probability distribution is infinitely divisible, therefore we would expect that our $C_j$ share this property. However, for a uniform distribution in the interval $(0,K)$ which is not infinitely divisible, the resulting $C_j$ are strictly positive, in fact $C_j = 2/(K+1)$, which invalidates the above statement (or, at least, weakens it considerably). The true origin of the oscillations still remains not fully specified.

Let us finally note that multiplicity distributions $P(N)$ are usually studied by analyzing factorial moments
\begin{equation}
F_q = \sum_{N=q}^{\infty} N(N-1)(N-2)\dots(N-q+1)P(N), \label{factmom}
\end{equation}
cumulant factorial moments,
\begin{equation}
K_q = F_q - \sum_{i=1}^{q-1}\binom{q-1}{i-1} K_{q-i}F_i \label{cumfactmom}
\end{equation}
or their ratios \cite{DH,Book-BP},
\begin{equation}
H_q = \frac{K_q}{F_q}, \label{RatioKqFq}
\end{equation}
which are very sensitive to the details of the multiplicity distribution. The advantage of their use is that they seem to be well described by perturbative QCD considerations, especially their oscillations in sign as a function of the rank $q$ \cite{DH}. On the other hand, $K_q$ can be expressed as an infinite series of the modified combinants, $C_j$ and, conversely, $C_j$ can be expressed in terms of $K_q$ \cite{Book-BP},
\begin{equation}
K_q = \sum_{j=q}^{\infty}\frac{(j-1)!}{(j-q)!}\langle N\rangle C_{j-1}\quad {\rm and}\quad C_j = \frac{1}{\langle N\rangle} \frac{1}{(j-1)!} \sum_{p=0}^{\infty}\frac{(-1)^p}{p!}K_{p+j}. \label{Connec}
\end{equation}
Note that the combinants  can be, by analogy to factorial cumulants, understood as {\it exclusive} correlation integrals \cite{Kittel,VVP}. However, they differ in the region of phase space they are most suitable to study: whereas cumulants are particularly well suited for the study of densely populated phase-space bins, combinants are better suited for the study of sparsely populated regions and their calculation requires only a finite number of $P(N)$, with $N<j$, which compensates the drawback caused by the requirement that one must have $P(0)>0$. Additionally, the advantage of combinants is that, being finite combinations of the probability ratios $P(N)/P(0)$, they do not suffer from a bias (empty-bin effect) present at high resolution in factorial moments and cumulants \cite{Kittel}. Combinants  share with cumulants the property of additivity, i.e., for a random variable composed of independent random variables, with its generating function given by the product of their generating functions, $G(x)=\prod_jG_j(x)$, the corresponding combinants are given by the sum of the independent components \cite{Book-BP}. Because the $C_j$ are directly connected with the $C^{\star}_j$ (cf. Eq. (\ref{connection})) they also share all their properties mentioned above. However, whether they share their oscillating pattern is still to be checked.

To summarize, we argue that only compound distributions based on the BD (like) and the NBD (like) components can fit adequately observed oscillations of modified combinants $C_j$. The question of which particular theoretical mechanism is at work remains, however, still open and one may expect a number of particular models to emerge here.

\medskip
Acknowledgements: This research  was supported in part by the National Science Center (NCN) under contracts 2016/23/B/ST2/00692 (MR) and 2016/22/M/ST2/00176 (GW).  We would like to thank Dr Nicholas Keeley for reading the manuscript.

\end{document}